\documentclass[pra,showpacs,twocolumn]{revtex4}
\usepackage{graphicx}
\usepackage{dcolumn}
\usepackage{bm}
\usepackage{amssymb}
\usepackage{amsmath}
\usepackage{epsfig}
\usepackage{times}

\begin{document}

\title{Mechanism of unidirectional emission of ultrahigh Q Whispering
Gallery mode in microcavities}
\author{C.-L. Zou}
\author{F.-W. Sun}
\email{fwsun@ustc.edu.cn}
\author{C.-H. Dong}
\author{X.-W. Wu}
\author{J.-M. Cui}
\author{Y. Yang}
\author{G.-C. Guo}
\author{Z.-F. Han}
\email{zfhan@ustc.edu.cn}
\affiliation{Key Lab of Quantum Information, University of Science and Technology of
China, Hefei 230026 }
\date{\today }

\begin{abstract}
The mechanism of unidirectional emission of high Q Whispering Gallery mode
in deformed circular micro-cavities is studied and firstly presented in this
paper. In phase space, light in the chaotic sea leaks out the cavity through
the refraction regions, whose positions are controlled by stable islands.
Moreover, the positions of fixed points according to the critical line in
unstable manifolds mainly determines the light leak out from which
refraction region and the direction of the emission. By controlling the
cavity shape, we can tune the leaky regions, as well as the positions of
fixed points, to achieve unidirectional emission high Q cavities with narrow
angular divergence both in high and low refractive index materials.
Especially for high index material, almost all Gibbous-shaped cavitiess have
unidirectional emission.
\end{abstract}

\maketitle

In recent years, being the high potential elements for integrated devices,
dielectric microcavities have attracted more and more attentions. Especial
attentions are focused on the Whispering Gallery (WG) micro-resonators, such
as micro-sphere, micro-disk, micro-toroid, etc, where energy can be well
confined in these rotational symmetrical structures by total internal
reflections. The unique high quality (Q) factor and low mode volume features
of WG Modes lead to wild applications, ranging from nonlinear optics, low
threshold lasers, sensitive sensors, to cavity quantum electrodynamics \cite%
{vahala}. However, the isotropic emission property of WGMs makes very low
efficient coupling from free space. Also, the near field coupling component
is a big obstacle for practical applications. To solve the coupling problem,
researchers found that the microcavity with slightly deformed circular
boundary, also called asymmetric resonant cavity (ARC), can lead directional
emission. The ARCs, such as the extensively studied quadruple \cite%
{qua1,qua2,qua3,qua4,qua5} and stadium microcavities \cite{sta1,sta2,sta3},
can be efficiently pumping through free space focused beam, which makes them
have been well used in the realization of strong coupling \cite{pump1} and
laser \cite{pump2}. Besides the easy free space coupling in the optical
experiments, the deformed microcavities can also serve as open billiards for
experimental research on quantum chaos \cite{chaos, chaos1}.

In order to get more efficient free space excitation and collection in
practical application, people are pursuing single directional emission with
narrow angular divergence, such as the spiral shaped \cite%
{chaos1,sprial1,spiral4} and the rounded isosceles triangle shaped \cite%
{triangle} microcavities. However, their low Q factors highly limit the
applications in low threshold laser and cavity quantum electronic dynamics.
Although there are several theoretical approaches \cite{annular,coupledisk},
the trade-off between the high Q and unidirectional emission still blocks
its further development. It is till recently that fabrication-friendly lima%
\c{c}on-shaped microcavity with both unidirectional emission and high Q was
theoretically proposed \cite{limacon} and experimentally realized by several
groups \cite{limaconexp1,limaconexp2,limaconexp3}. However, the mechanism to
design such cavity is not clear yet. Moreover, this lima\c{c}on-shaped
microcavity is based on the high refractive index material. There is no
report for high Q and unidirectional emission cavity with low index material.

In this paper, we will show the mechanism to get unidirectional emission
high Q mode both in the high and low refractive index ARC. At the beginning,
we summarize the necessary conditions to achieve the high Q unidirectional
emission ARC. First, the cavity boundary should be continuous, smooth, and
slightly deformed from a circle to support the WG-like modes. In such
deformed microcavity, the wave is mainly localized high above the critical
line and well confined in the cavity to support high Q. The light may leak
out when the ray is lower than critical line in phase space. In this case,
the rays follow the unstable manifolds, cross the critical line, and perform
the directional emission tangential to the cavity boundary.

Second, the boundary shape should have less symmetry. For planar deformed
cavity with two or more axial symmetries, there are at least two pairs of
symmetric tangential emission spots on the boundary because they support
both clockwise and counterclockwise traveling modes \cite%
{qua1,qua2,qua3,qua4,qua5,sta1,sta2,sta3}. The only way to achieve
unidirectional emission is to design the cavity with at most one symmetric
axis and make the emission direction along the axis direction.

In the cavity with continuous, smooth, and slightly deformed circular
boundary, there are stable and unstable orbits, corresponding to the stable
islands and fixed points of unstable manifolds in phase space. The mechanism
for the unidirectional emission in the single-axis cavity is: The light only
along the manifolds may leak out the cavity. Isolated by the stable islands,
the manifolds have several leaky regions when they cross the critical
refractive line. By controlling the cavity shape, we can tune the leaky
regions, as well as the position of fixed points according to the critical
line. The light through the lower fixed point has higher probability to leak
out than through the higher point. The parallel leaky lights along the axis
direction shows unidirectional emission from the high Q ARC. Based on this
mechanism, we succeed in designing the unidirectional high Q ARC both in
high and low refractive index materials. We find that almost all
Gibbous-shape ARCs have unidirectional emission for high refractive index
material. By precisely controlling the boundary shape, we can achieve far
field emission with very narrow divergence angle.

\begin{figure}[t]
\includegraphics [width=8cm] {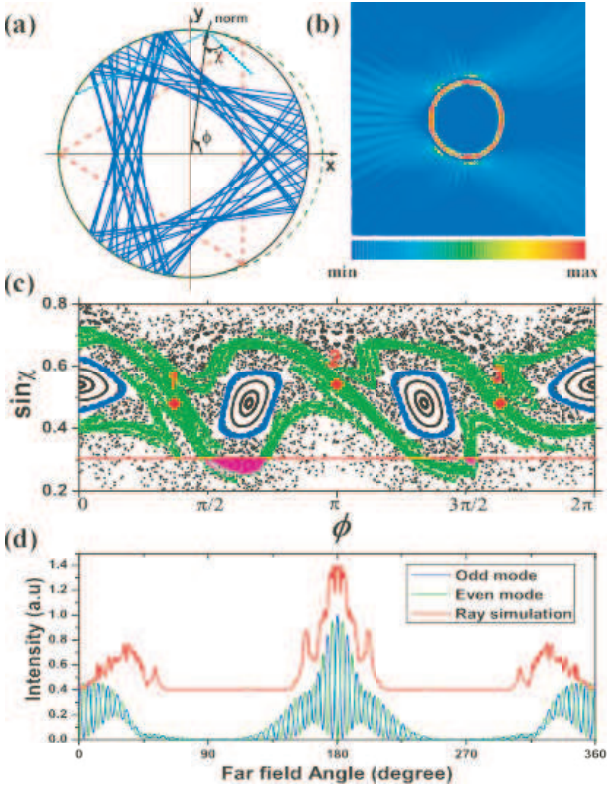}
\caption{(color online) (a) The real space illustration of the rays
reflection inside HQHC deformed microcavity with $\protect\epsilon =0.11$.
The dashed line outside is the circle boundary. The blue lines and red dash
lines are the stable and unstable period-3 orbits. (b) The calculated field
spatial distribution of TM polarized fundamental WG-like mode in HQHC, with
the wavenumber in vacuum $kr\approx 26.15$ and quality factor $Q=3.3\times
10^{5}$. (c) The phase space structure of HQHC. The blue circles and red
diamonds are the stable islands and fixed points, corresponding to the
stable and unstable period-3 orbits in (a). The red line is the critical
refraction line with $\sin \protect\chi =1/n$, and the green points are the
unstable manifolds from the unstable period-3 orbits. Magenta points is the
collection records of leaky rays location from the original rays upon $\sin
x>0.7$. It shows that most rays leak out from this refraction region. (d)
the far field patterns for TM polarized ray (the red line, lifted by $0.4$)
and wave simulation (blue line and green line for even and odd parity mode
respectively).}
\label{fig:epsart}
\end{figure}

At first, we consider a boundary shape of Half-Quadruple-Half-Circle (HQHC)
to illustrate the unidirectional emission in high refractive index cavity,
which have been demonstrated in silica microsphere \cite{hqhc}. The slightly
deformed circular boundary shape, as shown in Fig. 1(a), can be read as

\begin{equation}
R(\phi )=\left\{
\begin{array}{c}
R_{0},\text{\ \ } \\
R_{0}\left( 1-\epsilon \text{cos}^{2}\phi \right) ,%
\end{array}%
\text{\ }%
\begin{array}{c}
\cos \phi <0\text{,} \\
\cos \phi \geq 0\text{,}%
\end{array}%
\text{\ }\right.
\end{equation}%
where $\phi $ is the polar angle according to the symmetric X-axis and $%
\epsilon $ is the deformation factor. In the limit of $R\gg \lambda $, the
light in cavity can be semiclassically treated as ray. Although wave nature
of light is important in microcavity, the ray dynamics is still an efficient
tool to understand and analyze the emission properties of ARCs \cite%
{ray1,ray2,ray3,ray4,um1,um2,limaconexp2,limaconexp3}.

The sequence of ray reflection inside the cavity in real space can be
represented in the phase space as Poincar\'{e} Surface of Section (SOS).
Each ray reflection on the boundary is recorded in Birkhoff coordinates \cite%
{reichl} as $(\phi ,\sin \chi )$, where $\phi $ and $\chi $ denote the polar
angle of reflection position on boundary and the incidence angle,
respectively. By setting rays initially random on boundary (here we only
consider the counterclockwise propagation rays, the clockwise SOS can be
obtained by symmetric transformation), we can get the SOS by\ treating the
cavity as billiard \cite{chaos}, which is shown in Fig. 1(c).

In SOS, the phase space structure can be divided into two parts, the chaotic
sea and the stable islands.\ The two regions are isolated to each other. The
ray in chaotic sea [black points in Fig. 1(c)] has chaotic motions in real
space, while the ray in islands (in blue) is conserved and will never move
into chaotic sea. In the real space, the stable islands are corresponding to
the stable (blue solid lines) period-3 orbits \cite{reichl}, as shown in
Fig. 1(a). The light in the chaotic sea will run along the unstable
manifolds \cite{ray2} [green curve in Fig. 1(c)] and leak out when it
crosses critical line, which is denoted by the red line in Fig. 1(c). The
critical line shows the critical incidence angle for total internal
reflection $\sin \chi =1/n$. We set the refractive index $n=3.3$ in our case.

In this slightly deformed circular cavity, the light is mainly localized
upon the critical line in phase space. In the real space, the light runs in
the unstable periodic orbits or high Q WG-like mode, as shown in Fig. 1(b)
\cite{ray1,limacon,limaconexp1}. Initially, we set rays uniformly
distributed in the region $\sin \chi >0.7$, which can be considered as
energy of high Q modes dynamics localized above the critical line. Rays will
be reflected on the boundary in sequence until they hit the refraction
regions. In the SOS, the refraction regions are the parts of chaotic sea
below the critical line ($\sin \chi <1/n$). As the manifolds enclose the
stable islands, the locations of the stable islands determine the positions
of these refraction regions. In our case, the center two stable islands are
lower than the third island. So, there are two refraction regions near $\phi
\approx \pi /2$ and $3\pi /2$ at boundary, corresponding to two directional
emissions with far field angles of 180 and 360 degrees.

In order to get the unidirectional emission, we need to analyze the relative
positions of the three fixed points in the manifolds, as shown with red
diamonds in Fig.1(c). It is corresponding to the unstable (red dashed lines)
period-3 orbit \cite{reichl} in Fig. 1(a). They are sandwiched by the stable
islands. All lights following the manifolds will transit through/near these
three points. However, the fixed point (marked as $1$) before refraction
region near $\phi \approx \pi /2$ is lower than the point (marked as $2$)
before the other refraction region near $\phi \approx 3\pi /2$. Also, it is
closer to the refraction region. So the manifolds here are shorter and
steeper to cross the critical line.\textbf{\ }As a result, the light
through/near fixed point $1$ has the higher probability to enter the
fraction region and leak out than the other one. It will give rise to the
most energy leaky here and the directional emission with far field angle of
180 degrees. From the symmetry, the clockwise propagation rays leak out
around $\phi \approx 3\pi /2$ and show the same directional far field
emission. Therefore, the leaky beams from the counterclockwise and clockwise
propagations are parallel and finally show unidirectional emission from this
HQHC cavity.

With the numerical solution of Maxwell equations through boundary elements
method \cite{bem1, bem2}, we can observe the high Q WG-like mode in this
HQHC microcavity. Fig. 1(b) shows the electric field intensity distribution
of WG-like transverse magnetic (TM) polarized mode with radiation quantum
number $q=1$ (the transverse electric polarized modes have the similar
properties). The near field pattern shown tangent light emission from $\phi
\approx \pi /2$ and $3\pi /2$, corresponding to the two refraction regions.
Obviously directional emission beams for (counter)clockwise propagation are
along the X-axis.

The far field distribution of the Odd and Even polarity WG-mode is shown in
Fig. 2(d), as well as the ray-wave correspondence with ray dynamics. In the
ray simulation, with the weighted refractive coefficient from Fresnel law,
we can get the far field amplitude by summing up the all refractions, as
shown in Fig. 2(d) with the red curve. In order to get clear illustration,
the curve is lifted by 0.4. The far field pattern shows the unidirectional
emission at 180 degree with divergence about 30 degrees.

As discussed above, the refraction regions and fixed points in unstable
manifolds play the most important roles to determine the cavity directional
emission. So, we can tune the cavity shape to adjust the locations of stable
islands and the positions of fixed points to control the refraction regions
and the emission direction, simultaneously. To achieve the unidirectional
emission of high Q modes, we can design the X-axis symmetric cavity and lead
the unidirectional emission along the axis. In general, the boundary of
X-axis symmetric shape can be expressed as

\begin{equation}
R(\phi )=\left\{
\begin{array}{c}
R_{0}\sum a_{i}\cos ^{i}\phi , \\
R_{0}\sum b_{i}\cos ^{i}\phi ,%
\end{array}%
\text{\ }%
\begin{array}{c}
\text{ }\cos \phi \geq 0\text{\ \ } \\
\cos \phi <0%
\end{array}%
\text{\ }\right.
\end{equation}%
By setting $a_{0}=b_{0}=1$, and $a_{1}=b_{1}=0$, the norm direction is
continuous and the cavity boundary is smooth. In addition, we should keep
the boundary slow varying, so simply we cut off the high order terms, only
keep $a_{2}(b_{2})$ and $a_{3}(b_{3})$ nonzero \cite{note}. Also, to break
Y-axis symmetry, it needs $a_{2}\neq b_{2}$ or/and $a_{3}\neq -b_{3}$.
Moreover, we can set $b_{2}=b_{3}=0$ and $a_{2}+a_{3}<0$ to form Gibbous
shape for further simplification.

For the this kind of Gibbous shape cavity with high refractive index
material, there will always exist stable and unstable period-3 orbits with
vertexes near ($0,2\pi /3,4\pi /3$) and ($\pi /3,\pi ,5\pi /3$),
respectively. By randomly set the parameters $a_{i}$, we can always get
similar phase space structures to Fig. 1(c) in any Gibbous-shaped cavity.
That is because, without Y-axis symmetric, the two sets of period-3 orbits
in Gibbous shape cavities always have smaller convex angles near $\pi /2$ in
real space. Correspondingly, the fixed points of manifolds and stable
islands are lower in phase space, which make the refraction regions fixed at
about $\phi \approx \pi /2$ and $3\pi /2$ and unidirectional emission from
the first refraction region. We carried out ray and wave simulations on lots
of Gibbous shape cavities. Results indicate the single emission with far
field divergence angle ranges from $20$ to $50$ degrees, confirming to our
conjecture from the orbits. We find the cavity with $%
a_{2}=0.0486,a_{3}=-0.1258$ gives good unidirectional far field pattern. The
wave simulation gives a high Q ($Q=5.78\times 10^{6}$) TM polarized WG mode
with $kr\approx 25.1842$, and the far field divergence is only about $24$
degrees, which is much smaller, comparing to $40$ degrees reported in lima\c{%
c}on cavity \cite{limacon}.

Thus, we have shown how to get unidirectional emission in the Gibbous-shaped
cavity. The lima\c{c}on cavity \cite{limacon}, as well as the cavity in
Song's experiments \cite{limaconexp1}, is a generalized Gibbous-shaped
cavity. Their mechanisms for unidirectional emission can also be explained
with our approach.

\begin{figure}[t]
\includegraphics [width=8cm] {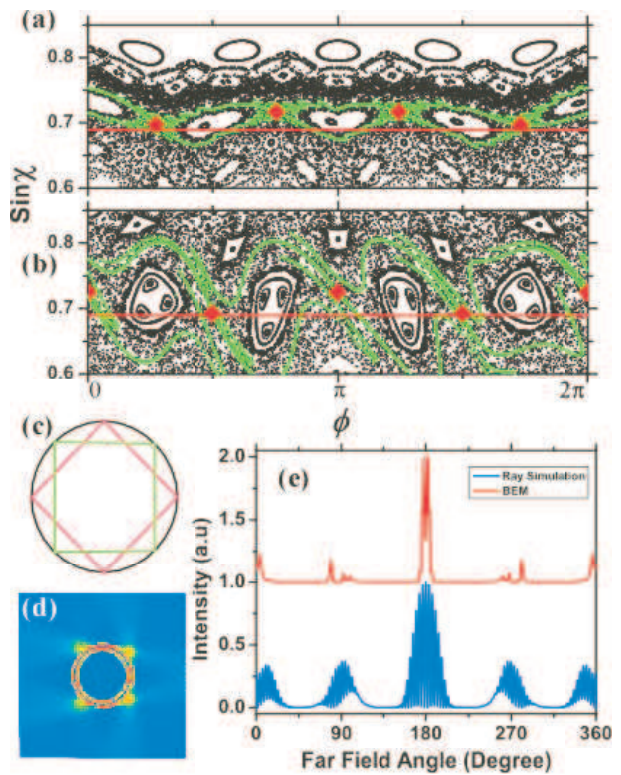}
\caption{(color online)(a) (b) the SOS of two cavities (a) HQHC $\protect\epsilon =0.05$
and (b) the cavity shape support stable rectangle orbit and unidirectional
emission designed in the text. (c) the rectangle and diamond shape period-4
orbits in the X-symmetric cavity. (d) The near field distribution of the
unidirectional emission with $kr\approx 52.5284$. (e) The far field pattern
of the cavity, red line is the ray simulation result (lifted by 1), and the
blue line is corresponding to the result in (d).}
\end{figure}

Now, we will generalize the above mechanism for unidirectional emission to
low refractive index microcavities. Here we take the silica ($n=1.45$) for
example. Similar to the period-3 orbits in high index cavities, the silica
cavity's directional emission is basically influenced by period-4 orbits.
The former experimental study on HQHC micro-sphere showed that the emission
direction could be reduced to 2 nearly perpendicular direction \cite{hqhc}.
The SOS of such HQHC is present in Fig. 2(a), where the stable period-4
islands [the red diamond orbit in Fig.2(c)] around the critical line prevent
the rays unidirectional emission around $\phi =\pi /2$.

A strategy to solve the problem in this low index cavity is to find a shape
with stable islands away from $\pi /2$. That is to make the green rectangle
period-4 orbit be stable and the red diamond orbit be unstable.
Corresponding SOS is presented in Fig. 2(b). Rays can go cross the critical
line near $\pi /2$. Similar to the analysis in high index cavity, the
conditions for unidirectional emission in silica microcavity is: (i) The
cavity supports the stable rectangle orbit. (ii) The stable island near $%
\phi =3\pi /4$ and $5\pi /4$ is lower, and the fixed point (the red triangle
in SOS) is lower near $\phi =\pi /2$.

In the numerical simulation of ray dynamics, we randomly choose parameters
to set the boundary shape, judge that the rectangle orbit stable or not, and
compare the islands position in SOS to fulfill the condition (ii). We can
easily find a good cavity shape with $%
a_{2}=-0.1329,a_{3}=0.0948,b_{2}=-0.0642,b_{3}=-0.0224$ satisfies the above
conditions, and gives good unidirectional far field pattern. Fig. 2(b) and
Fig. 2(d) shows the SOS and field distribution of TM mode. Fig.2(e) is the
corresponding far field pattern compares to the ray simulation. As we
expect, the Q factor is up to $2.75\times 10^{6}$ and divergence is only
about 30 degree.

Thanks to the ray-wave correspondence, an efficient route to design a
unidirectional emission cavity shape could be: using the ray dynamics to
give the SOS of the selective boundary shape, predicting the emission
properties, and carrying out the wave simulation to confirm it. Here, we
only gives examples on the typically materials, such as semiconductor and
silicon ($n\approx 3.3$) and silica glass ($n\approx 1.45$). Other popular
solid materials for photonics have similar refractive index. There are also
many materials not at this range, such as some doped glass with the
refractive index around $2.0$. We have also successfully designed the cavity
shapes to achieve high Q and unidirectional emission (not presented here).
By adjusting the periodic-3 orbits or periodic-4 orbits, we can get
appropriate emission direction and expect the unidirectional emission in the
materials with refractive index range from $1.4$ to $4$.

There are also some deviations between the simulations with ray and wave,
especially when $kr$ is small. The actions of light could not totally be
described by rays. Better ray-wave correspondence could be expected in
larger cavity \cite{limaconexp3}. In smaller cavity, wave properties should
be included, such as, the diffractions and interference. In microcavities,
researches have illustrated that some modifications should be included in
the ray dynamics \cite{wave}, such as the corrected Fresnel's law at curve
interface \cite{fresnel}, the Goos-H\"{a}nchen shift \cite{gh} and the
Fresnel Filtering effect \cite{filt}.

In conclusion, we examined and presented the mechanism for high Q
unidirectional emission WG modes in micro-cavities. Based on the necessary
condition for the continuous and single axis symmetric boundary shape, we
can well control the directional emission from ARC by setting appropriate
boundary shape to adjust the stable islands and fixed points of unstable
orbits in phase space. With the assistance of ray dynamics, it is easy to
design fabrication-friendly simple cavity boundary shape to achieve high Q
and unidirectional emission in different materials with the refractive index
range from $1.4$ to $4$. We expect our approaches presented here to discuss
the light in cavities can be applied to the study of the chaotic transport
in two dimensional phase space \cite{Shim08}, as well as other opening
non-integrable systems, such as quantum dots and nano-structures.

\begin{acknowledgments}
C.-L. Zou thanks Juhee Yang and J. Wiersig for discussions. Financial
support by the National fundamental Research Program of China under Grant No
2006CB921900; National Science Foundation of China under Grant No. 60537020
and 60621064; The Knowledge Innovation Project of Chinese Academy of
Sciences \& Chinese Academy of Sciences and International Partnership
Project; Y. Yang is also funded by the China Postdoctoral Science
Foundation. F.-W. Sun is also supported by the starting funds from USTC for
new faculty.
\end{acknowledgments}


\begin{thebibliography}{99}
\bibitem{note} It is not required for general cases. Here we just set the
values of high orders zero for simplicity in simulation, as well as
simplification in fabrication.
\end{thebibliography}
\end{document}